\documentclass[conference]{IEEEtran}
\usepackage[noadjust]{cite}
\usepackage{makeidx}  
\usepackage{amsmath}
\usepackage{amsfonts}
\usepackage{cite}
\usepackage{wrapfig}
\usepackage{graphicx}
\usepackage{color}
\usepackage{moresize}
\graphicspath{ {figures/} }
\ifCLASSINFOpdf
\else
\fi
\begin{document}
%

\title{Towards tailoring non-invasive brain stimulation using real-time fMRI and Bayesian optimization}

\author{
\IEEEauthorblockN{Romy Lorenz\IEEEauthorrefmark{1}\IEEEauthorrefmark{2}\IEEEauthorrefmark{3},
Ricardo P Monti\IEEEauthorrefmark{1}\IEEEauthorrefmark{4}, 
Adam Hampshire\IEEEauthorrefmark{2},
Yury Koush\IEEEauthorrefmark{5},
Christoforos Anagnostopoulos\IEEEauthorrefmark{4},\\ 
Aldo A Faisal\IEEEauthorrefmark{3}, 
David Sharp\IEEEauthorrefmark{2}, 
Giovanni Montana\IEEEauthorrefmark{4}\IEEEauthorrefmark{6}, 
Robert Leech\IEEEauthorrefmark{2} and 
Ines R Violante\IEEEauthorrefmark{2}}
\\
\IEEEauthorblockA{\IEEEauthorrefmark{2}Department of Medicine, Imperial College London, UK, }
\IEEEauthorblockA{\IEEEauthorrefmark{3}Department of Bioengineering, Imperial College London, UK}
\IEEEauthorblockA{\IEEEauthorrefmark{4}Department of Mathematics, Imperial College London, UK}
\IEEEauthorblockA{\IEEEauthorrefmark{5}Institute of Bioengineering, EPFL, Switzerland}
\IEEEauthorblockA{\IEEEauthorrefmark{6}Department of Biomedical Engineering, King{'}s College London, UK}
\IEEEauthorblockA{\IEEEauthorrefmark{1} \emph{These authors contributed equally to this work.}}
}

\author{
\IEEEauthorblockN{Romy Lorenz\IEEEauthorrefmark{1}\IEEEauthorrefmark{2},
Ricardo P Monti\IEEEauthorrefmark{1}\IEEEauthorrefmark{2}, 
Adam Hampshire\IEEEauthorrefmark{2},
Yury Koush\IEEEauthorrefmark{3},
Christoforos Anagnostopoulos\IEEEauthorrefmark{2},\\ 
Aldo A Faisal\IEEEauthorrefmark{2}, 
David Sharp\IEEEauthorrefmark{2}, 
Giovanni Montana\IEEEauthorrefmark{2}\IEEEauthorrefmark{4}, 
Robert Leech\IEEEauthorrefmark{2} and 
Ines R Violante\IEEEauthorrefmark{2}}
\\
\IEEEauthorblockA{\IEEEauthorrefmark{2}Imperial College London, UK, \IEEEauthorrefmark{3}EPFL, Switzerland,  \IEEEauthorrefmark{4}King{'}s College London, UK}
\IEEEauthorblockA{\IEEEauthorrefmark{1} \small{\emph{These authors contributed equally to this work.}}}
}


\maketitle

\begin{abstract}
Non-invasive brain stimulation, such as transcranial alternating current stimulation (tACS) provides a powerful tool to directly modulate brain oscillations that mediate complex cognitive processes. While the body of evidence about the effect of tACS on behavioral and cognitive performance is constantly growing, those studies fail to address the importance of subject-specific stimulation protocols. With this study here, we set the foundation to combine tACS with a recently presented framework that utilizes real-time fRMI and Bayesian optimization in order to identify the most optimal tACS protocol for a given individual. While Bayesian optimization is particularly relevant to such a scenario, its success depends on two fundamental choices: the choice of covariance kernel for the Gaussian process prior as well as the choice of acquisition function that guides the search. Using empirical (functional neuroimaging) as well as simulation data, we identified the squared exponential kernel and the upper confidence bound acquisition function to work best for our problem. These results will be used to inform our upcoming real-time experiments. 
\end{abstract}


%
\IEEEpeerreviewmaketitle

\section{Introduction}
\begin{figure}
 	\centering
      \includegraphics[scale=0.5]{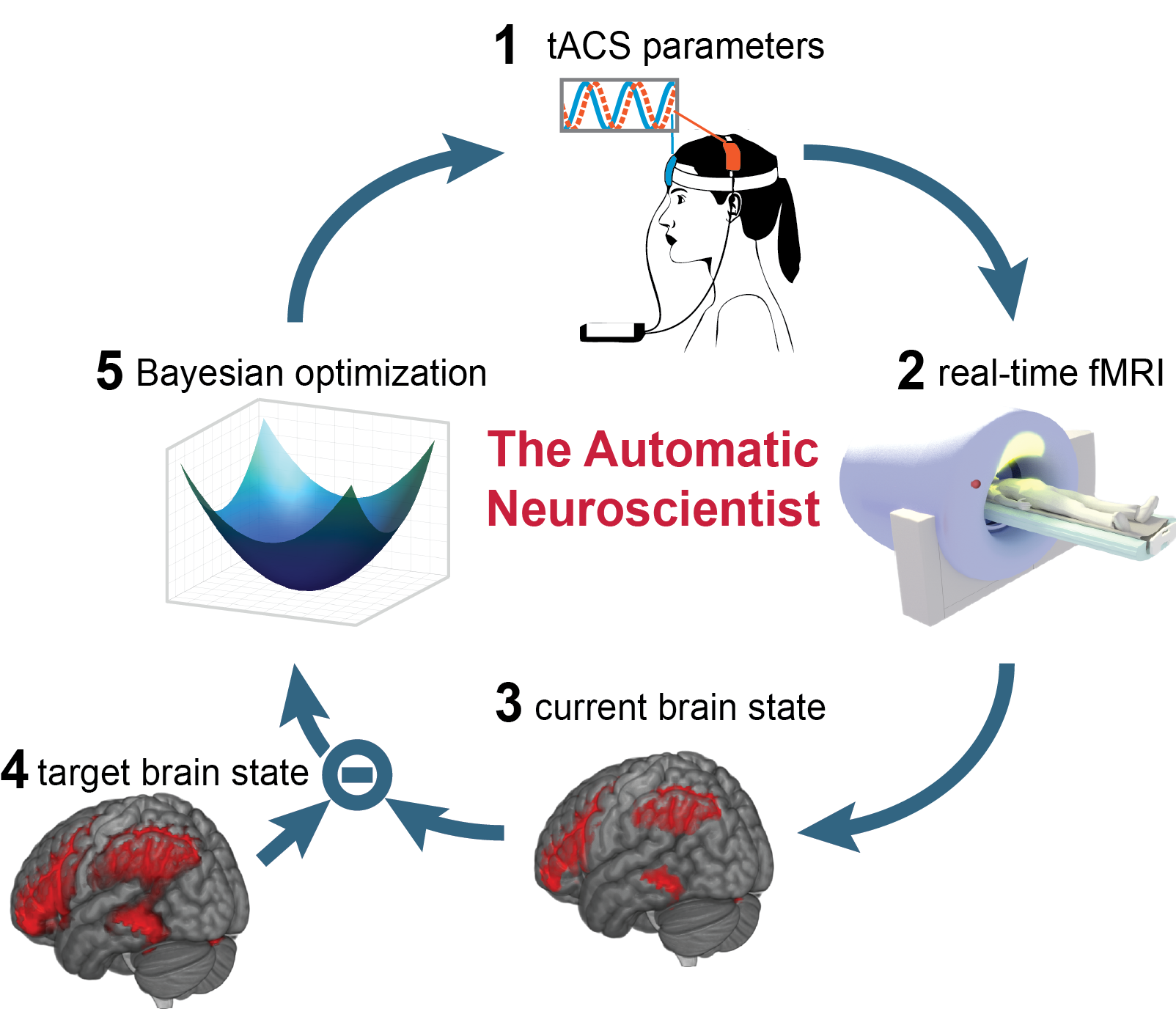}
  \caption{High-level overview of the \emph{Automatic Neuroscientist} combined with tACS. \textbf{(1)} The experiment starts with applying a  random combination of tACS parameters to the subject in the scanner. \textbf{(2)} Whole-brain functional images are acquired and analyzed in real-time in response to the block of stimulation. \textbf{(3)} Information about the current brain state is extracted and \textbf{(4)} compared to the pre-defined target brain state. This result is then fed into the Bayesian optimization approach. \textbf{(5)} Based on this, the algorithm chooses a new combination of tACS parameters that optimizes for the target brain state. This closed-loop cycle continues until the optimal tACS parameters are found. }
  \label{fig:AutomaticNeuroscientist}
\end{figure}

Studies involving non-invasive brain stimulation have reported remarkable changes in cognitive and behavioral performance \cite{Kuo}. Among those techniques, transcranial alternating current stimulation (tACS) is particularly promising as it directly allows for the  modulation of physiologically relevant brain oscillations that subserve cognitive operations \cite{Herrmann}. The effects of tACS are highly dependent on the frequency and phase of stimulation \cite{Antal}. 
The conventional tACS approach involves defining the frequency and phase of stimulation ad hoc and testing them on a cohort of subjects. However, this approach exhibits two limitations: (1) the brain networks targeted by the stimulation cannot be verified without simultaneous fMRI; (2) those stimulation parameters may vary across subjects due to difference in anatomy or heterogeneity in disease. Yet, there is a combinatorial explosion in the biologically plausible range of stimulation frequencies (1-100 Hz) and phases (0-359$^\circ$), resulting in up to thousands of possibilities. Identifying the optimal stimulation protocol for a given individual is like ``finding a needle in a haystack". Therefore, using conventional methodology makes tailoring tACS to an individual highly unfeasible. To address this fundamental challenge in the clinical use of tACS, we aim to combine tACS with a recently presented framework that behavior real-time functional magnetic resonance imaging (fMRI) and Bayesian optimization: The \emph{Automatic Neuroscientist} \cite{Lorenz}. Employing such a framework allows us to start with a target brain state and find a set of tACS parameters (frequency and phase of stimulation) for each individual that maximally activates it (Fig.~\ref{fig:AutomaticNeuroscientist}).

Bayesian optimization is particularly relevant to such a problem as it provides a powerful strategy for finding maxima of objective functions that are expensive to evaluate and might contain noisy measurements (both criteria are met with functional neuroimaging data). Moreover, Bayesian optimization is suited in situations where we do not have an analytical expression of the objective function nor can make formal statements regarding its properties \cite{Brochu,Shahriari}. However, the implementation of Bayesian optimization requires several fundamental choices that are paramount to the success of the technique; such as (1) the choice of covariance function for the Gaussian process (GP) prior and (2) the choice of acquisition function 
which determines the manner in which the space of parameters is explored.  

The aim of this study is to objectively compare the performance of a variety of distinct kernel functions (resulting in distinct GP priors) and acquisitions functions.
Offline fMRI data from eight healthy volunteers receiving blocks of non-invasive tACS stimulation over left frontoparietal brain regions as well as simulation data were employed.  The results obtained will be used to propose a combination of kernel and acquisition functions for upcoming real-time experiments.



\section{Methods}

The objective of this work is to obtain a 
better understanding of the 
relationship between various combinations of tACS parameters (i.e., frequency and phase) and each subject's neural response. 
In the framework of Bayesian optimization, such a relationship is summarized by a latent objective function which we wish 
to infer. 
With this in mind, a small neuroimaging study was conducted where healthy participants received tACS
consisting of various combinations of frequency and phase. 
This data was subsequently employed to perform model selection on the GP covariance function. In addition to this, simulation analyses were carried out to 
compare two popular acquisition functions.


\subsection{Empirical Data}

\subsubsection{Subjects and Experimental Design}
\begin{figure}
 	\centering
      \includegraphics[scale=0.5]{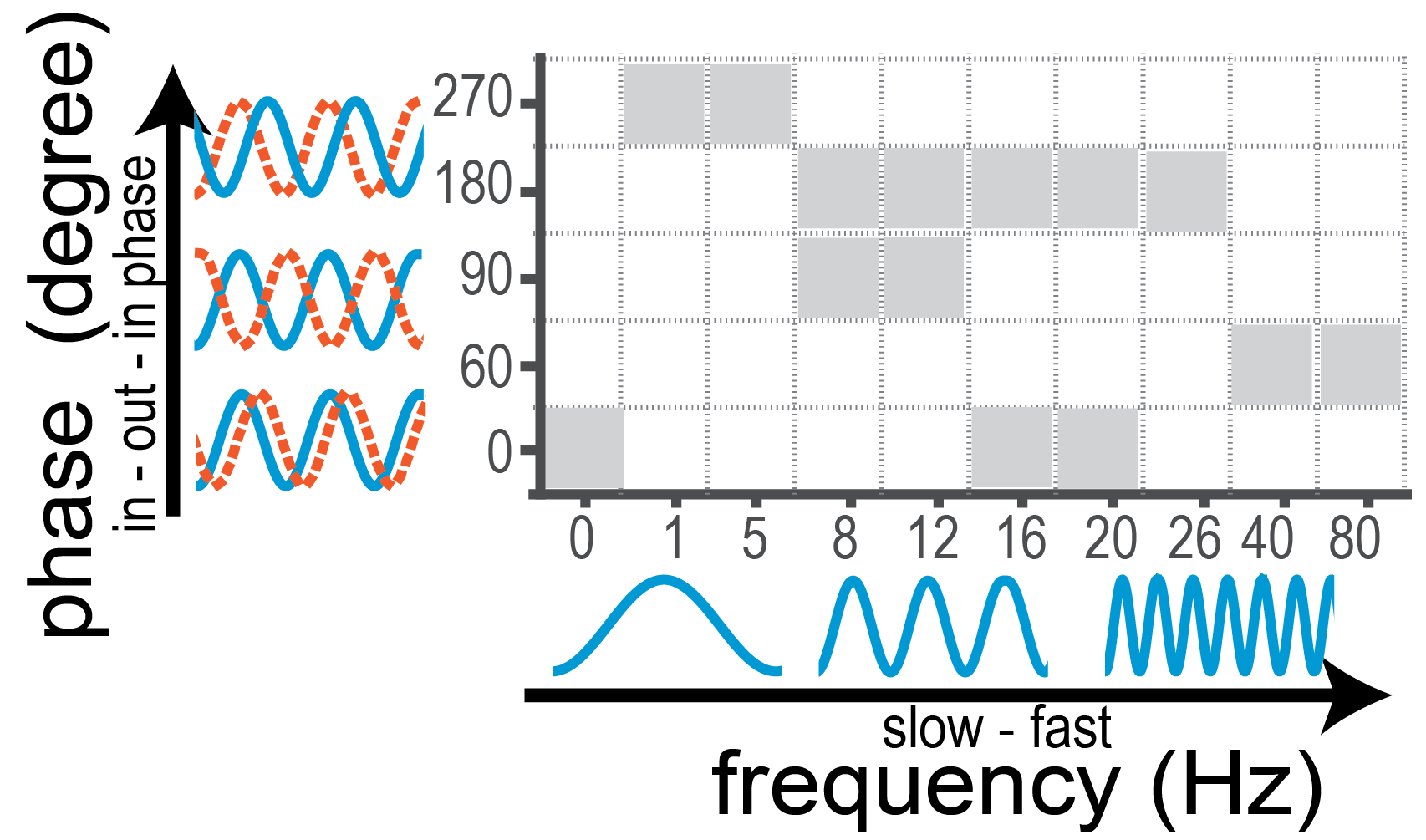}
  \caption{Exhaustive two-dimensional tACS parameter space with each dimension corresponding to the frequency or phase of non-invasive brain stimulation, respectively. Grey shaded squares are combinations sampled in our (offline) fMRI study.}
  \label{fig:Space}
\end{figure}

Eight healthy volunteers (5 female, 24.75 $\pm$ 3.49 years) took part in our study. The study was approved by the Hammersmith Hospital (London, UK) research ethics committee. Each participant performed three runs of a switch task \cite{Dove} in the MR scanner. Each run consisted of task blocks (36 s) interleaved with rest blocks (24 s). During task blocks, tACS was applied via a pair of MR-compatible conductive rubber electrodes over left frontal (F3) and left parietal (P3) regions (as determined by the International 10-20 EEG system). Both return electrodes were placed on the ipsilateral shoulder. The current was fixed to 1 mA (peak-to peak). The experiment parameter space was limited to ten different frequencies (0,1,5,8,12,16,20,26,40 and 80 Hz) and five different phases (0, 60, 90, 180 and 270$^\circ$), resulting in 50 different frequency-and-phase combinations (see Fig.~\ref{fig:Space}). For each run and each subject, we tested a subset of 14 identical frequency-and-phase combinations (grey shaded squares in Fig.~\ref{fig:Space}). The order of those frequency-and-phase combinations was random for each run in order to eliminate any order effect of the stimulation on the subject's neural response. The condition without any stimulation (0 Hz with 0$^\circ$) was the only condition that was repeated two (n=4) to four (n=4) times within each run. 

\subsubsection{Target brain state extraction}
Whole-brain coverage images were acquired by a Siemens Verio 3T scanner using an EPI sequence (TR: 1.5s). Data was minimally preprocessed by applying motion-correction, high-pass filtering (100s) and spatial smoothing (5mm FWHM Gaussian kernel). We than extracted 20 regions of interest (ROIs) that have been previously shown to predict trial-by-trial performance in task switching (for ROI definition we created 8mm spheres around the peak coordinates reported in \cite{Leber}). The extracted timecourses were further cleaned by removing high-frequency noise and large signal spikes using a modified Kalman filter \cite{Koush}. We then assessed changes in functional connectivity between those regions for blocks of tACS using psychophysiological interaction (PPI) analyses \cite{Friston} (whilst also including six motion parameters as confounds). For the analyses presented here, we selected the two regions that exhibited the highest inter-run robustness across all subjects (assessed using {Spearman}'s rank correlation coefficient): the left inferior parietal lobule and the posterior cingulate gyrus. In order to assess how likely the pattern of results between those regions could have occurred by chance, we performed non-parametric permutation testing (10,000 permutations) and found the result to be highly significant (\emph{z}=3.71, \emph{p}$<$.001).

\subsection{Bayesian optimization}
The underlying intuition behind the method of Lorenz et al. \cite{Lorenz} is that the target brain pattern is a function of experimental conditions. As such, the authors propose to learn the relationship by modelling the observed brain state as a sample from a Gaussian process (GP). This facilitates the use of a Bayesian optimization framework \cite{Brochu,Shahriari} in a closed-loop form where subjects are presented with an experimental condition and real-time fMRI provides instantaneous information about the subject's brain state. Based on this, we can iteratively update our beliefs about the unknown objective function, captured in the posterior distribution of the GP. While Lorenz et al. \cite{Lorenz} used simple perceptual stimuli in their original experiment, our study here involves blocks of non-invasive brain stimulation with different combinations of frequency and phase. This can be considered as a far more challenging problem as only a handful of studies exist investigating the effects of tACS on the blood oxygenation level dependent (BOLD). Beyond that, high inter-subject variability can be expected as outlined in the Introduction. 

In contrast to the fMRI setting employed in \cite{Lorenz}, there is far less prior knowledge regarding the properties of the latent objective function which can be 
leveraged in this context. Therefore, data was collected over a cohort of eight subjects (see previous section). Using this empirical data we begin by studying the choice of kernel function in Section \ref{Covariance_function}. Given a choice of kernel and by carrying out simulation analyses, the effects of various acquisition functions are subsequently assessed in Section \ref{Acquisition_function}.



\subsubsection{Covariance function}
\label{Covariance_function}

The choice of kernel function in GP regression is fundamental \cite{GPbook}. The kernel directly specifies a
measure of similarity across various inputs and therefore determines the  generalization properties of the 
estimated model. 
Model selection in  the context of GPs is often posed as a parameter estimation task
whereby the hyper-parameters for a specific choice of kernel can be learnt \cite{GPbook}.
However, 
comparing the performance of potentially many distinct kernels is a challenging problem. 
Recently, \cite{Duvenaud, lloyd2014automatic} have proposed a method for 
automatically searching over the space of kernel structures in a principled manner. 

Briefly, the proposed approach proceeds by considering 
compositional structures (either addition or multiplication) over four 
\textit{base kernels}; squared exponential (as employed in \cite{Lorenz}),
periodic, linear or rational quadratic. 
When comparing various distinct kernel structures the 
Bayesian Information Criterion (BIC) is employed 
to score the various models (where type-2 maximum likelihood estimates of parameters are used). 
Initially, 
each of the base kernels is proposed. Thereafter, the selected kernel can be extended by either 
performing addition  or multiplication with any base kernel. 
While the original approach described in \cite{Duvenaud} proposes kernels for each dimension independently, 
in this work both dimensions were studied simultaneously (thus two dimensional kernels were proposed at
each step). This decision is based on the hypothesis that each dimension (either frequency or phase) 
would share a similar relationship with the response, therefore leading to the selection of similar (if not the same)
kernel.

\subsubsection{Acquisition function}
\label{Acquisition_function}

Of equal importance for the success of the Bayesian optimization is the choice of acquisition function. 
The role of the acquisition function is to iteratively propose 
the experimental condition to be presented to the subject. It therefore serves to guide 
the search over all combinations of parameters (in our case frequency and phase) and must 
implicitly balance exploration with exploitation.
In this work we considered the performance of two popular acquisition functions:
the expected improvement (EI) and upper confidence bound (GP-UCB) acquisition function \cite{Brochu}\footnote{
The probability of improvement (PI) acquisition 
function was not included as it is typically seen to be over exploitative.}. The two acquisition functions differ in their trade-off between exploration and exploitation of the experimental parameters space, giving rise to distinct sampling behaviors over time. Informally, the EI acquisition function can be seen as trying to maximize the expected improvement over the current best. While the GP-UCB also favors the selection of points with high mean value (similar to the EI acquisition function), it also favors points with high variance (i.e., regions worth to explore). In certain settings, this will result in a more explorative behavior of the GP-UCB compared to the more ``greedy" EI acquisition function. For algorithmic details, please refer to \cite{Brochu}.

While the choice of kernel function can be guided by model selection and information theoretic measures,
the comparison of multiple acquisition functions requires explicit knowledge of the underlying objective function.
As a result, a simulation study was employed to the benchmark the performance of the two aforementioned 
acquisition functions.

\begin{figure*} [t]
 	\centering
      \includegraphics[width=\textwidth]{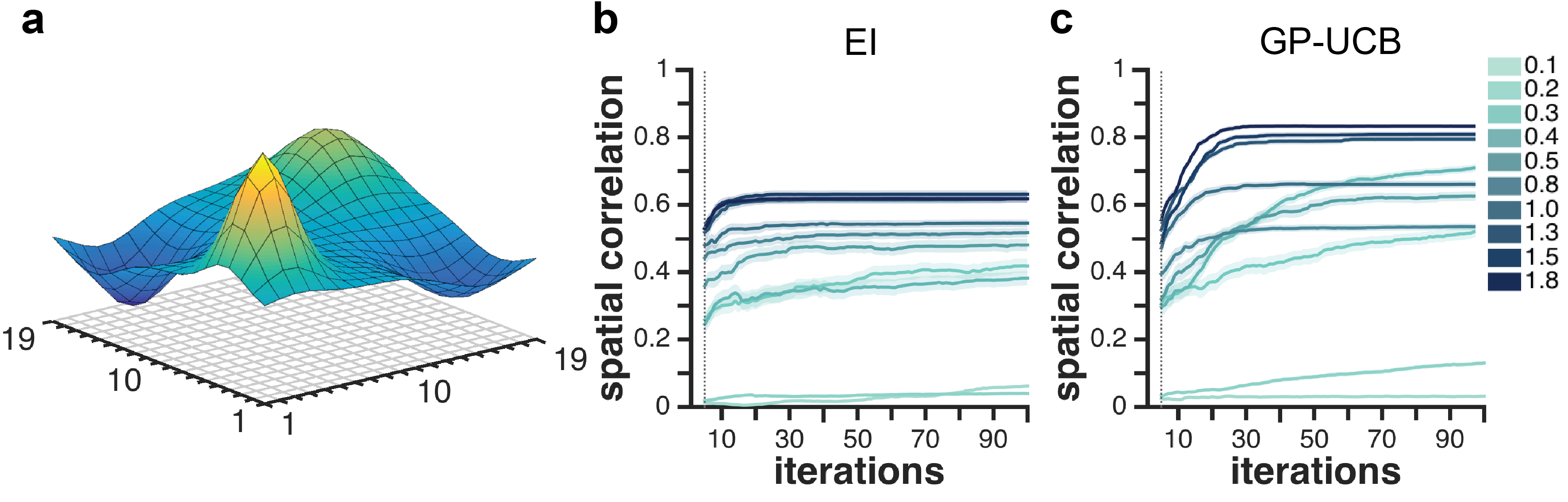}
  \caption{Results of simulation analyses. \textbf{(a)} Modeled objective function used for simulations. Mean $\pm$  SEM (shaded areas) spatial correlation between predicted and modeled parameter space for \textbf{(b)} EI  and \textbf{(c)} GP-UCB acquisition function. As the first five iterations were used as a burn-in, they are not depicted here (gray dashed line). Simulations were performed for 10 different CNR values, ranging from 0.1 (bright blue) to 1.8 (dark blue).}
  \label{fig:Simulations}
\end{figure*}

A complex and multimodal objective function  was proposed 
for the simulation study (see Fig.~\ref{fig:Simulations}a). This was motivated by our empirical results, showing multiple optima and minima for each subject (not shown).  As the presence of non-neural noise is well documented for fMRI experiments, we also studied how different levels of contrast-to-noise-ratio (CNR) affected the Bayesian optimization. We repeated our simulations analyses with CNR values ranging from 0.1 and 1.8 (typically reported CNR values in fMRI literature vary between 0.5 and 1.8 \cite{Welvaert}). For each CNR value tested, we ran 100 simulations. The maximum number of iterations was set to 100 with the first five (randomly selected) observations serving as burn-in phase. Thereafter, at each iteration, a new frequency-and-phase combination was proposed by maximizing the respective acquisition function and sampled by the Bayesian optimization algorithm in the next iteration. 

As a measure of accuracy we computed spatial correlation between the algorithm's predictions for the whole parameter space and the ``ground truth" objective function. Considering our underlying motivation to gain a holistic understanding of the whole parameter space (learning about maxima and minima), spatial correlations seemed most appropriate to capture similarity across the whole space. This procedure was identical to the one reported in \cite{Lorenz}.


\section{Results}


\subsection{Covariance kernel selection}

In this work we follow \cite{Duvenaud, lloyd2014automatic}, and look 
to compare several distinct kernels via the use of BIC. Furthermore, we also employ 
a greedy search algorithm whereby we begin with a base kernel and 
iteratively extend this kernel. 

There are two significant differences in the approaches described in this work 
to those of \cite{Duvenaud, lloyd2014automatic}. The first is our choice of base kernels;
in this work the squared exponential, periodic, linear and Matern $(\frac{3}{2})$ where 
employed as base kernels. The second difference is that two-dimensional kernels were proposed 
at each iteration. 
This choice was motivated by an assumption that the relationship between parameter (frequency or phase) and the response 
shared a similar functional form. 
Finally, a minor difference between this work and that of \cite{lloyd2014automatic} is that 
change-point functions were not considered here. 

Following \cite{Duvenaud}, we begin by proposing each of the base kernels and calculating the associated BIC scores.
In this first step the squared exponential kernel was selected. We then proceeded to consider 
compositional structures. Each of these structures was composed by either adding or multiplying the squared exponential 
kernel with each of the base kernels. For each compositional kernel, the parameters were estimated using the previous 
parameters as warm starts. Again, BIC scores were employed to score each of the compositional models.
At this stage, there was no reduction in BIC scores resulting in the choice of the squared exponential kernel.

\subsection{Acquisition function selection}

Simulation analyses were carried out to assess the performance of the EI and GP-UCB acquisition functions. 
A range of CRN values were employed in order to recreate the properties of fMRI data as well as
study the performance of each of the acquisition functions in scenarios with low signal-to-noise ratio. 
Throughout this simulation the underlying GP model was maintained constant, allowing us to attribute 
any differences in performance to the acquisition functions. 

For each CNR value, 100 simulations where performed. Each simulation allowed the 
proposed Bayesian optimization algorithm to explore the parameter space over 100 iterations. The 
spatial correlation between the predictive posterior under the GP model and the true objective function 
was calculated at each iteration. 

The mean $\pm$  SEM spatial correlation results are depicted in Fig.~\ref{fig:Simulations}b for the EI and in Fig.~\ref{fig:Simulations}c for the GP-UCB acquisition function. 
Not surprisingly, we note that both acquisition functions fail in the presence of too much noise (for CNR values $<0.2$).
The results also indicate that the 
GP-UCB acquisition function clearly outperforms the EI acquisition function for typical CNR values reported in the literature (0.5-1.8).

\section{Discussion and future work}
The results presented here serve as an initial exploration 
into the choices of kernel and acquisition functions which can be used in the context 
of real-time optimization for tACS. 
Our results propose the use of a squared exponential kernel function in combination with a GP-UCB acquisition 
function. In a series of simulations, the latter was shown to 
capture the whole parameter space (as identified by using spatial correlations) when compared to the EI acquisition function.
This result was consistent for a range of CNR values tested. These findings serve as an important basis for upcoming real-time experiments. An avenue of future work is the definition of appropriate stopping criteria to determine convergence \cite{Lorenz2}. This is particularly important in our context due to high scanning costs, limited attentional capacities of subjects and neural habituation to the stimulation.


To our knowledge, this is the first study to propose a framework that allows tailoring non-invasive brain stimulation to an individual. The next step consists of applying this technology to patients suffering from traumatic brain injury; a condition that disrupts frontoparietal brain networks and hence results in cognitive impairments \cite{Jilka}. While closed-loop deep brain stimulation has already been shown to outperform the conventional approach in {Parkinson}'s patients \cite{Little} we envision that our framework will advance personalized treatment by means of non-invasive brain stimulation. This will be of particular importance for neurological and psychiatric deficits that are diffuse and widely heterogeneous in their origin.

\section*{Acknowledgment}
This research was supported by the NIHR Imperial BRC and the Wellcome Trust (103045/Z/13/Z).




\begin{thebibliography}{1}

\scriptsize{
\bibitem{Kuo}
MF Kuo, MA Nitsche: Effects of transcranial electrical stimulation on cognition, Clin. EEG Neurosci.,43(3), 92--199 (2012)}

\bibitem{Herrmann}
CS Herrmann, S Rach, T Neuling, D Strueber: Transcranial alternating current stimulation: a review of the underlying mechanisms and modulation of cognitive processes, Frontiers Human Neuroscience, 7:279 (2013)

\bibitem{Antal}
A Antal, W Paulus: Transcranial alternating current stimulation (tACS), Frontiers Human Neuroscience, 7:317 (2013)

\bibitem{Lorenz}
R Lorenz, RP Monti, IR Violante et al.: The Automatic Neuroscientist: A framework for optimizing experimental design with closed-loop real-time fMRI. NeuroImage,129, 320--334 (2016)

\bibitem{Brochu}
E Brochu, VM Cora, N de Freitas: 
A tutorial on Bayesian optimization of expensive cost functions, with application to active user modeling and hierarchical reinforcement learning.
arXiv:1012.2599 (2010)

\bibitem{Shahriari}
B Shahriari, K Swersky, Z Wang, RP Adams, N de Freitas: Taking the Human Out of the Loop: A Review of Bayesian Optimization, Proceedings of the IEEE, 104(1), 148--175 (2016)

\bibitem{Dove}
A Dove, S Pollmann, T Schubert, CJ Wiggins, DY von Cramon: Prefrontal cortex activation in task switching: an event-related fMRI study. Cognitive Brain Research, 9(1): 103--109 (2000)

\bibitem{Leber}
AB Leber, NB Turk-Browne, MM Chun: Neural predictors of moment-to-moment fluctuations in cognitive flexibility, PNAS, 105(36), 13592--13597 (2008)

\bibitem{Koush}
Y Koush, M Zvyagintsev, M Dyck, KA Mathiak, K Mathiak: Signal quality and {Bayesian} signal processing in neurofeedback based on real-time {fMRI}, NeuroImage, 59(1) (2012)

\bibitem{Friston}
KJ Friston: Functional and Effective Connectivity: A Review, Brain Connectivity, 1(1),13--36 (2011)


\bibitem{GPbook}
CE Rasmussen, CK Williams:  Gaussian processes for machine learning. MIT Press (2006)

\bibitem{Duvenaud}
D Duvenaud, JR Lloyd, R Grosse, JB Tenenbaum, Z Ghahramani.: 
Structure discovery in nonparametric regression through compositional kernel search,
ICML (2013)

\bibitem{lloyd2014automatic}
JT Lloyd, D Duvenaud, R Grosse, JB Tenenbaum, Z Ghahramani.:
Automatic construction and natural-language description of nonparametric regression models, Association for the Advancement of Artificial Intelligence 
(2014)

\bibitem{Welvaert} 
M Welvaert, Y Rosseel: On the definition of signal-to-noise ratio and contrast-to-noise ratio for fMRI data, PLoS One, 8, doi: 10.1371/journal.pone.0077089 (2013)

\bibitem{Lorenz2}
R Lorenz, RP Monti, IR Violante et al.: Stopping criteria for boosting automatic experimental design using real-time fMRI with Bayesian optimization, arXiv:1511.07827 (2015)


\bibitem{Jilka} 
SR Jilka, G Scott, T Ham, A Pickering, V Bonnelle, RM Braga, R Leech, DJ Sharp: Damage to the Salience Network and Interactions with the Default Mode Network, J. Neurosci., 34(33), 0798-10807 (2014)

\bibitem{Little}
S Little, A Pogosyan, S Neal et al.:Adaptive deep brain stimulation in advanced Parkinson disease,  Ann. Neurol., 74(3), 449--457 (2013)

\end{thebibliography}
%

\bibliographystyle{IEEEtran}

\scriptsize{

}

\end{document}